\documentclass[useAMS,usegraphicx,usenatbib]{mn2e}          

\usepackage{times}
\usepackage{flushend}
\usepackage{amsmath,amssymb}

\newcommand{\Msun}{\ensuremath{\,{\rm M}_\odot}}                  
\newcommand{\Rsun}{\ensuremath{\,{\rm R}_\odot}}                  
\newcommand{\Teff}{\ensuremath{T_{\rm eff}}}                      
\newcommand{\kms}{\,km\,s$^{-1}$}                                 

\newcommand{\reff}[1]{#1}

\title[$\beta$\,Cep pulsations in V453\,Cyg]
      {Discovery of $\beta$\,Cep pulsations in the eclipsing binary V453\,Cygni}

\author[Southworth et al.]
       {John Southworth\,$^{1}$, D.\ M.\ Bowman\,$^2$, A.\ Tkachenko\,$^2$, K.\ Pavlovski$^3$ \\
        $^1$\,Astrophysics Group, Keele University, Staffordshire, ST5 5BG, UK \\
        $^2$\,Institute of Astronomy, KU Leuven, Celestijnenlaan 200D, B-3001 Leuven, Belgium \\
        $^3$\,Department of Physics, Faculty of Science, University of Zagreb, Bijenicka cesta 32, 10000 Zagreb, Croatia
        }

\begin{document} \maketitle 

\begin{abstract}
V453\,Cyg is an eclipsing binary containing 14\Msun\ and 11\Msun\ stars in an eccentric short-period orbit. We have discovered $\beta$\,Cep-type pulsations in this system using TESS data. We identify \reff{seven} significant pulsation frequencies, between \reff{2.37} and 10.51\,d$^{-1}$, in the primary star. These include six frequencies which are separated by yet significantly offset from harmonics of the orbital frequency, indicating they are tidally-perturbed modes. We have determined the physical properties of the system to high precision: V453\,Cyg\,A is the first $\beta$\,Cep pulsator with a precise mass measurement. The system is a vital tracer of the physical processes that govern the evolution of massive single and binary stars.
\end{abstract}



\section{Introduction}
\label{sec:intro}

Massive stars -- those with initial masses $\ga$8\Msun\ -- are important drivers in the chemical and dynamical evolution of galaxies \citep{Langer12araa}. Despite their importance in the cosmos, we still lack prescriptions for their interior rotation, mixing and angular momentum transport mechanisms \citep{Aerts++19araa,Aerts20rmp}. Since massive stars harbour convective cores during the hydrogen-core burning phase of stellar evolution, mixing in the near-core region directly supplies fresh hydrogen to the core and prolongs the main-sequence lifetime. However, the amount of near-core mixing within massive stars is observationally unconstrained, which makes it challenging to accurately and reliably infer the absolute ages of single stars from only their location in the HR diagram. The physical correlations and degeneracies between model parameters in current stellar evolution codes are more important for massive stars in binary systems since many of these will interact during their lifetimes \citep{Sana+12sci}.

One of the best methods for breaking model degeneracies is by using eclipsing binary stars (EBs), as their masses and radii can be measured entirely empirically and to high precision, and because we can assume that the stars have the same age and chemical composition \citep{Torres++10aarv}. The physical properties can be compared to the predictions of theoretical evolutionary models to infer the age and composition of the components, and to investigate the effectiveness of different parameterisations of the interior physics inplemented in the codes. However, there is often significant disagreement between the masses of O and B stars inferred from spectroscopy and stellar evolution theory, versus dynamical masses measurements in binary systems. This \textit{mass discrepancy} was identified by \citet{Herrero+92aa} and has been explained by invoking varying amounts of near-core mixing \citep{Guinan+00apj,Tkachenko+14mn,Tkachenko+16mn,Tkachenko+20xxx}.

\begin{figure*} \includegraphics[width=\textwidth]{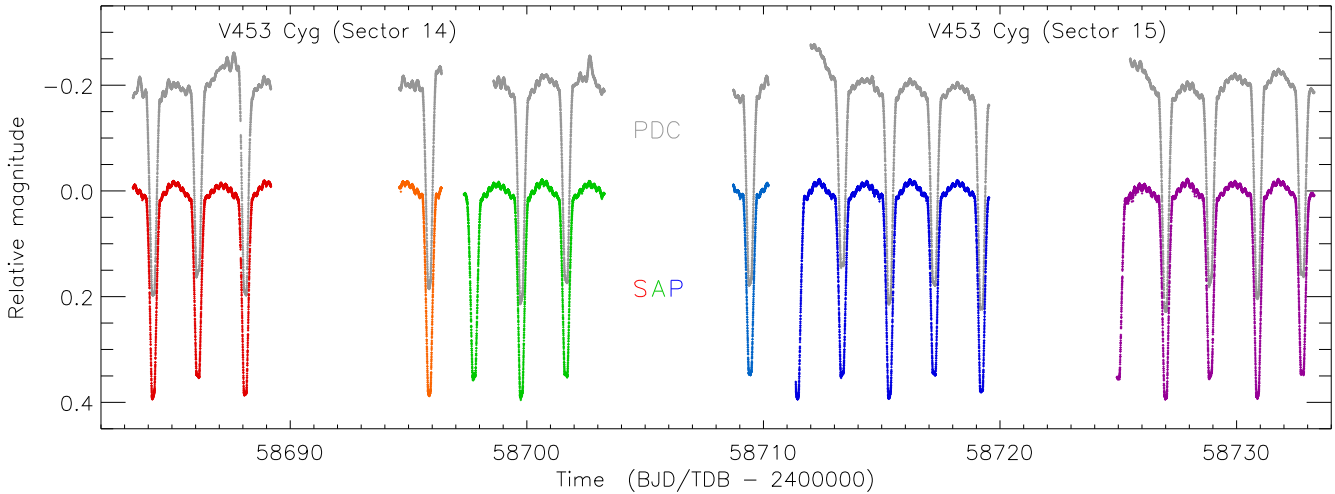}
\caption{\label{fig:v453:tess} TESS light curve of V453\,Cyg. The PDC data are
shown in grey and the SAP data are shown in six colours. The colours indicate
the division of the data into six sets for the polynomial normalisation
functions.} \end{figure*}

Amongst massive stars are a type of pulsating variable known as $\beta$\,Cep stars, which have masses 8--15\Msun\ \citep{Aerts++10book}. These stars pulsate in low-radial order gravity (g) and pressure (p) modes with pulsation periods of several hours and pulsation amplitudes up to a few tenths of a magnitude \citep{StankovHandler05apjs}. Pulsations have great potential in providing constraints on stellar interiors, since each pulsation mode is sensitive to a specific pulsation cavity within a star. For example, the identification of rotationally-split p-modes has allowed the interior rotation to be determined \citep{Aerts+03sci,Briquet+07mn} and the impact of magnetic fields to be constrained \citep{ShibahashiAerts00apj,Briquet+13aa}. The launch of space \reff{missions} such as CoRoT, \textit{Kepler} and K2 has allowed many new pulsating massive stars to be discovered \citep[e.g.][]{Neiner+12aa,McNamara++12aj,Papics+17aa,Burssens+19mn}.

A unique opportunity exists in the study of pulsating stars in binary systems. Here, the measured masses and radii fix the positions of the stars in the mass--radius and HR diagrams, whilst the pulsations allow a direct probe of interior rotation and mixing. Furthermore, pulsating binary systems have the potential to constrain the effect of tides on stellar structure and evolution by means of tidally-induced pulsations in eccentric systems \citep{Welsh+11apjs,Hambleton+13mn,Beck+14aa,Fuller17mn,Handler+20natas}, and the recently-discovered tidally-perturbed pulsations in short-period binaries \citep{Bowman+19apj}. However, no precise mass measurement currently exists for a $\beta$\,Cep pulsator. A few have been found in binary systems, for example EN\,Lac \citep{Jerzykiewicz+15mn}, V381\,Car \citep{Freyhammer+05aa} and $\alpha$\,Vir \citep{Tkachenko+16mn} but in each case the mass is not precisely known because either the secondary star is not detectable in spectra or the orbital inclination is poorly determined due to the lack of eclipses.

In this work we present the discovery of $\beta$\,Cep pulsations in V453\,Cyg, an EB consisting of B0.4\,IV + B0.7\,IV stars. Its 3.89\,d orbit is slightly eccentric and exhibits fast apsidal motion  \citep{Wachmann73aa}. It is a member of NGC\,6871, a sparse open cluster that is the nucleus of Cyg\,OB3 \citep{Hoag+61pusno,GarmanyStencel92aas}. V453\,Cyg was discovered to be eclipsing by \citet{Wachmann39bzan}, and light curves have been published by \citet{Wachmann74aa}, \citet{Cohen74aas} and \citet{Zakirov92kfnt}. Spectroscopic orbits have been obtained by \citet{PopperHill91aj}, \citet{SimonSturm94aa} and \citet{Burkholder++97apj}, chemical abundances by \citet{PavlovskiMe09mn} and \citet{Pavlovski++18mn}, and physical properties by \citet{Me++04mn2} and \citet{Pavlovski++18mn}.


\section{Observations}                                                                                                              \label{sec:obs}

V453\,Cyg was observed using the NASA Transiting Exoplanet Survey Satellite (TESS; \citealt{Ricker+15jatis}). TESS is observing the majority of the sky, with each hemisphere divided up into 13 sectors based on ecliptic longitude. Observations in each sector last for 27\,d, with an interruption for data download near the midpoint.


V453\,Cyg was observed in Sectors 14 and 15, between dates 2019/07/18 and 2019/09/11, at a cadence of 2\,min. Fig.\,\ref{fig:v453:tess} shows the SAP (simple aperture photometry) and PDC (Pre-search Data Conditioning) light curves \citep{Jenkins+16spie}.


We applied the least-squares deconvolution technique \citep{Tkachenko+13aa} to two published sets of spectra \citep{Me++04mn2,Pavlovski++18mn} and find clear signs of line profile variations in the primary star. We therefore attribute the pulsations to this star, but caution that the data were not good enough to demonstrate that the secondary star is \emph{not} pulsating.


\section{Physical properties of V453\,Cyg}

\subsection{Preliminary light curve analysis: \textsc{jktebop}}
\label{sec:jktebop}

The TESS light curves of V453\,Cyg were initially modelled using the {\sc jktebop} code\footnote{\texttt{http://www.astro.keele.ac.uk/jkt/codes/jktebop.html}} \citep{Me08mn,Me13aa}. This is a fast code which provides a good fit to the eclipses, so is useful for normalising the data to zero differential magnitude and for isolating the pulsation signatures. However, it approximates the stars as spheres so the parameters of the fit are not reliable for stars as distorted as those in V453\,Cyg.

We fitted both the PDC and SAP light curves, but our final results rest on the SAP data as the systematic effects are smaller in this light curve; use of the PDC data has no significant effect on our conclusions. Due to the sometimes large systematics, we divided the SAP data into chunks of contiguous datapoints and included a polynomial function of second or third order to fit the out-of-eclipse brightness in each. There are 22\,473 SAP datapoints in our fits.

We fitted for the sum and ratio of the fractional radii of the stars ($r_{\rm A} = \frac{R_{\rm A}}{a}$ and $r_{\rm B} = \frac{R_{\rm B}}{a}$ where $R_{\rm A}$ and $R_{\rm B}$ are the true radii and $a$ is the orbital semimajor axis), the orbital inclination, the central surface brightness ratio, orbital period and a reference time of mid-transit. We did not use historical light curves or times of miniumum light in order to avoid the complexities induced by the apsidal motion of the system. Limb darkening was included using the quadratic law with the linear coefficients fitted and the quadratic coefficients fixed to values from \citet{Claret17aa}. The orbital eccentricity ($e$) and argument of periastron ($\omega$) were fitted using the Poincar\'e elements ($e\cos\omega$ and $e\sin\omega$). 

In order to account for the pulsations simultaneously with the effects of binarity, we added to {\sc jktebop} the option to fit for one or more sine terms applied to various parameters. We experimented with including one sine term applied to the light of the system or of either of the two stars. \reff{All three options yielded a significantly better fit to the data, but are} statistically indistinguishable so could not be used to confirm the spectroscopic result that the primary star is the pulsator in the system.

\begin{figure} \includegraphics[width=\columnwidth]{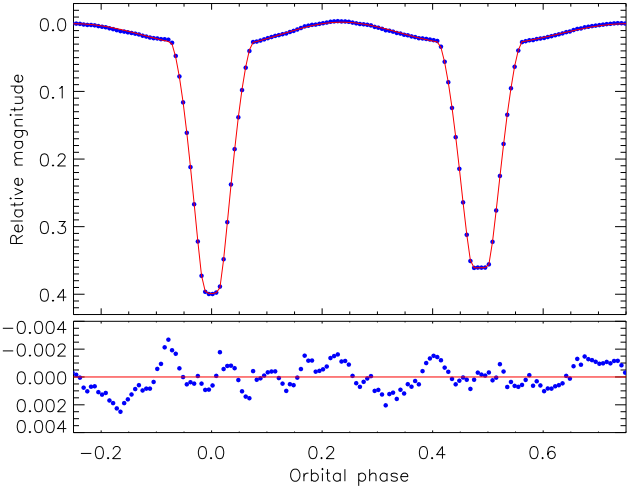}
\caption{\label{fig:v453:phase} Phase-binned SAP light curve of V453\,Cyg
(filled blue circles) with the {\sc wd2004} best fit (red solid line). The
residuals of the fit are shown on a large scale in the lower panel.} \end{figure}

\subsection{Final light curve analysis: Wilson-Devinney}

The TESS light curve was fitted using the Wilson-Devinney (WD) code \citep{WilsonDevinney71apj}, which uses Roche geometry to accurately model the observable properties of binary systems. We used the 2004 version of this code, driven via the {\sc jktwd} wrapper \citep{Me+11mn}. Due to the much longer computing time needed for the WD code, particularly in the case of an eccentric system, we modelled a light curve that had already been cleaned using {\sc jktebop}. This cleaning process comprised normalisation to zero differential magnitude, removal of the sinusoidal variation due to the strongest pulsation, and phase-binning into 300 datapoints.

This binned light curve was modelled with {\sc jktwd} using the same approach as adopted in \citet{Pavlovski++18mn}, with the one modification that we always fitted for the light output of the two stars separately because the TESS passband is not implemented in the 2004 version of the WD code. \reff{The control and fixed parameters of the final fit are given in Table\,\ref{tab:absdim}; see \citet{WilsonVanhamme04} for a detailed description of each quantity.} The Poisson noise in the light curve is negligible, so the uncertainties in the parameters derived from the data are dominated by choices made during the modelling process. In order to capture these uncertainties we ran solutions with different approaches to the treatment of eccentricity, third light, rotation rate, albedo, gravity darkening, treatment of limb darkening, and treatment of reflection. The adopted model and the uncertainties from the variety of models considered are collected in Table\,\ref{tab:absdim}. We found it important to include albedo as a fitted parameter, as this significantly decreased the residuals of the fit. The albedo values are positively correlated with the assumed gravity darkening exponents. The best fit is shown in Fig.\,\ref{fig:v453:phase}, where the residuals come primarily from incomplete removal of the pulsations \reff{at this stage}.

\begin{table}
\caption{\label{tab:absdim} Brief summary of the parameters for the WD solution of 
the TESS light curve of V453\,Cyg. Uncertainties are only quoted when they have 
been robustly assessed by comparison between a full set of alternative solutions.}
\setlength{\tabcolsep}{5pt}
\begin{tabular}{@{}lcc@{}} \hline
Parameter                                & Star A                & Star B                \\                
\hline                                                                                                     %
{\it Control parameters:} \\                                                                               %
{\sc wd2004} operation mode              & \multicolumn{2}{c}{0}                         \\                
Treatment of reflection                  & \multicolumn{2}{c}{1}                         \\                
Number of reflections                    & \multicolumn{2}{c}{1}                         \\                
Limb darkening law                       & \multicolumn{2}{c}{1 (linear)}                \\                
Numerical grid size (normal)             & \multicolumn{2}{c}{60}                        \\                
Numerical grid size (coarse)             & \multicolumn{2}{c}{40}                        \\[3pt]           
{\it Fixed parameters:} \\                                                                                 %
Mass ratio                               & \multicolumn{2}{c}{0.795}                     \\                
Rotation rates                           & 1.0                   & 1.4                   \\                
Gravity darkening                        & 1.0                   & 1.0                   \\                
\Teff\ (K)                               & 28\,800               & 27\,700               \\[3pt]           
{\it Fitted parameters:} \\                                                                                %
Orbital inclination (\degr)              & \multicolumn{2}{c}{$86.14 \pm 0.30$}          \\                
Orbital eccentricity                     & \multicolumn{2}{c}{$0.0250 \pm 0.0014$}       \\                
Argument of periastron (\degr)           & \multicolumn{2}{c}{$152.5 \pm 5.1$}           \\                
Third light                              & \multicolumn{2}{c}{$0.002 \pm 0.013$}         \\                
Light contributions                      & $9.079 \pm 0.020$     & $3.026 \pm 0.035$     \\                
Potential                                & $4.383 \pm 0.013$     & $5.762 \pm 0.043$     \\                
Bolometric albedos                       & $1.33 \pm 0.44$       & $1.07 \pm 0.35$       \\                
Fractional radii                         & $0.2844 \pm 0.0010$   & $0.1723 \pm 0.0016$   \\[3pt]           
{\it Derived parameters:} \\                                                                               %
Orbital separation (\Rsun)               & \multicolumn{2}{c}{$30.47 \pm 0.16$}          \\                
Mass (\Msun)                             & $13.96 \pm 0.23$      & $11.10 \pm 0.18$      \\                
Radius (\Rsun)                           & $8.665 \pm 0.055$     & $5.250 \pm 0.056$     \\                
Log surface gravity (cgs)                & $3.708 \pm 0.004$     & $4.044 \pm 0.009$     \\                
\Teff\ (K)                               & $28\,800 \pm 500$     & $27\,700 \pm 600$     \\                
Log luminosity (L$_\odot$)               & $4.666 \pm 0.031$     & $4.163 \pm 0.039$     \\                
Absolute bolometric magnitude            & $-6.914 \pm 0.077$    & $-5.657 \pm 0.097$    \\                
Distance (pc)                            & \multicolumn{2}{c}{$1851 \pm 60$}             \\                
\hline
\end{tabular}
\end{table}

\subsection{Physical properties of V453\,Cyg}

The light curve parameters found in the previous section were augmented \reff{by the velocity amplitudes ($K_{\rm A} = 175.2 \pm 1.3$\kms\ and $K_{\rm B} = 220.2 \pm 1.6$\kms) and effective temperatures (${\Teff}_{\rm A} = 28\,800 \pm 500$\,K and ${\Teff}_{\rm B} = 27\,700 \pm 600$\,K) of the stars. These were taken from \citet{Pavlovski++18mn}, who obtained them by disentangling the spectra \citep{SimonSturm94aa,Ilijic+04aspc} and fitting them with a grid of non-LTE synthetic spectra.} We then used the {\sc jktabsdim} code \citep{Me++05aa} to calculate the physical properties of the system, propagating the uncertainties using a perturbation analysis. The distance to the system was calculated from its $BVRI$ apparent magnitudes, the bolometric corrections from \citet{Girardi+02aa}, and by manual iteration of the reddening parameter $E_{B-V} = 0.44 \pm 0.03$ to bring the distances from different passbands into agreement.


The masses of the stars are measured to an accuracy of 1.6\% and their radii to 0.6\% and 1.0\%. The {\sc jktebop} code gives radii that differ from those found using {\sc wd2004} by 2\%. We interpret this as an indication that {\sc jktebop} does not give reliable parameters for stars as tidally distorted as those of V453\,Cyg. The two codes agree much better for stars with smaller fractional radii, e.g.\ to within 0.1\% when $r_{\rm A} = 0.04$ and $r_{\rm B} = 0.06$ \citep{Maxted+20mn}.

The \reff{pseudo}synchronous rotational velocities of the primary and secondary stars are $112.7 \pm 0.7$\kms\ and $68.3 \pm 0.7$\kms, respectively, whereas the $v\sin i$ values measured by \citet{Pavlovski++18mn} are $107.2 \pm 2.8$\kms\ and $98.3 \pm 3.7$\kms. We therefore agree with the conclusion of \citeauthor{Pavlovski++18mn} that the primary is rotating approximately pseudo-synchronously with the orbit whereas the secondary is rotating faster than this, by a factor of $1.44 \pm 0.08$.

\begin{figure*} \includegraphics[width=1.03\textwidth]{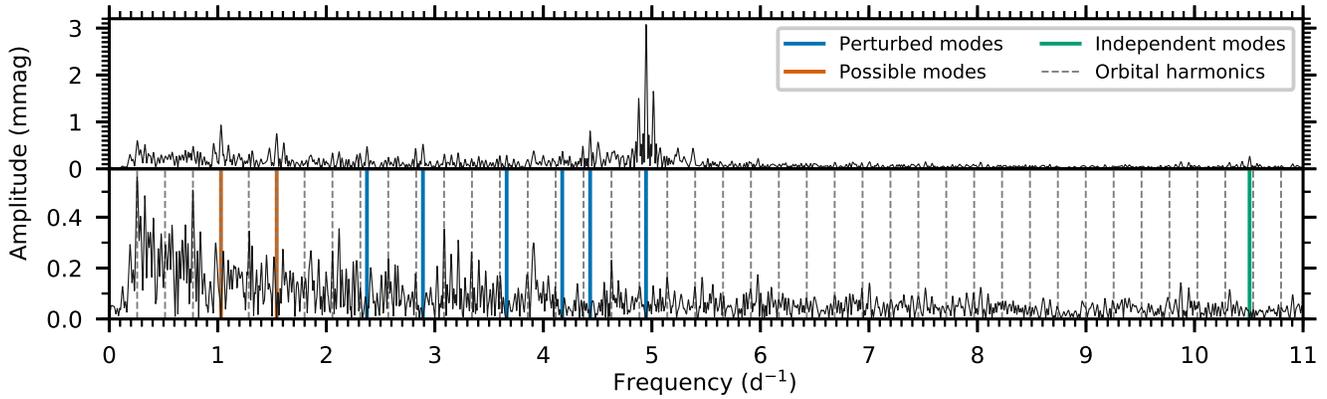}
\caption{\label{fig:v453:FT} Amplitude spectra of V453\,Cyg after removal of the binary model  
(top), and also the significant pulsation frequencies (bottom). 
The location and identity of extracted frequencies are shown as coloured vertical lines, 
and the vertical dashed-grey lines represent multiples of the orbital frequency.} \end{figure*}


\section{Pulsation analysis}

After subtracting the binary model from the TESS light curve (Sect.\,\ref{sec:jktebop}), we used a discrete Fourier transform (\citealt{Deeming75apss,Kurtz85mn}) to calculate the amplitude spectrum of the remaining variability in V453\,Cyg. The 2-min-cadence TESS data have a Nyquist frequency of 360\,d$^{-1}$. We extracted all significant pulsation modes, defined as having an amplitude signal-to-noise ratio (S/N) larger than four \citep{Breger93apss}, using iterative pre-whitening, then performed a multi-frequency non-linear least-squares fit to optimise their parameters \citep{Kurtz+15mn,Bowman17book}. The nine significant frequencies are given in Table\,\ref{tab:puls}.

\begin{table} \centering
\caption{\label{tab:puls}Frequencies, amplitudes and phases of significant frequencies in V453\,Cyg. The $1\sigma$ 
uncertainties from the multi-frequency non-linear least-squares fit are given in brackets. The amplitude S/N and 
the difference between each frequency and a harmonic, $i$, of the orbital frequency are given.}
\setlength{\tabcolsep}{4pt}
\begin{tabular}{rrrrrr}
\hline
\multicolumn{1}{c}{Frequency}  & \multicolumn{1}{c}{Amplitude} & \multicolumn{1}{c}{Phase} & \multicolumn{1}{c}{S/N} & \multicolumn{1}{c}{$i$} & \multicolumn{1}{c}{$\nu - i\nu_{\rm orb}$} \\
\multicolumn{1}{c}{(d$^{-1}$)} & \multicolumn{1}{c}{(mmag)}    & \multicolumn{1}{c}{(rad)} & \multicolumn{1}{c}{}    & \multicolumn{1}{c}{}    & \multicolumn{1}{c}{(d$^{-1}$)}             \\
\hline
$1.02900 \,(18)$ & $0.983\,(17)$ & -1.32\,(3)  & $ 5.2$ &  4 &  0.0007\,(8) \\
$1.54215 \,(23)$ & $0.766\,(17)$ & -1.20\,(4)  & $ 5.0$ &  6 & -0.0003\,(8) \\
$2.37468 \,(42)$ & $0.423\,(17)$ &  1.63\,(8)  & $ 4.0$ &  9 &  0.0610\,(9) \\
$2.88971 \,(35)$ & $0.504\,(17)$ & -1.11\,(7)  & $ 4.6$ & 11 &  0.0618\,(9) \\
$3.66118 \,(63)$ & $0.285\,(17)$ &  1.55\,(12) & $ 4.0$ & 14 &  0.0621\,(10) \\
$4.17402 \,(49)$ & $0.366\,(17)$ &  1.93\,(9)  & $ 4.1$ & 16 &  0.0607\,(10) \\
$4.43103 \,(19)$ & $0.924\,(17)$ &  0.23\,(4)  & $10.5$ & 17 &  0.0607\,(8) \\
$4.94605 \,( 6)$ & $3.099\,(17)$ &  0.52\,(1)  & $30.4$ & 19 &  0.0615\,(8) \\
$10.50733\,(68)$ & $0.262\,(17)$ & -1.55\,(13) & $ 6.8$ & 41 & -0.0330\,(11) \\
\hline
\end{tabular}
\end{table}

Two of the frequencies in Table~\ref{tab:puls}, $1.03$ and $1.54$~d$^{-1}$, are within 3.7$\sigma$ and 1.5$\sigma$ of the fourth and sixth harmonics of the orbital frequency, respectively. \reff{These frequencies could be tidally-induced modes, but, given the extremely high precision of the TESS light curve, we cannot exclude that they result from the subtraction of an imperfect binary model, which leaves residual signal at orbital harmonics.} The highest frequency, $10.51$~d$^{-1}$ is not a combination of any other frequency and likely an independent heat-driven p-mode. Assuming the $1\sigma$ confidence intervals in the masses and radii from our binary model, we estimate using pulsation constants and the period-mean density relation (see e.g.\ \citealt{Bowman17book}), that the fundamental p-mode frequency \reff{is} $4.07(7)$ and $7.7(2)$~d$^{-1}$ in the primary and secondary, respectively. Thus the extracted frequencies represent low-radial order g- and p-modes.


The TESS data alone do not allow us to firmly ascertain which star is pulsating. We do not find amplitude modulation of the pulsation modes during the orbit, so V453\,Cyg is not a single-sided pulsator \citep{Handler+20natas}. However, we do find regularity among six significant frequencies, which exhibit a quasi-constant offset from orbital-frequency harmonics and are indicative of tidally-perturbed pulsations \citep{Bowman+19apj}. The amplitude spectra before and after pre-whitening are shown in the top and bottom panels of Fig.\,\ref{fig:v453:FT}, respectively, in which the location of extracted frequencies are shown as colour-coded vertical lines. The vertical dashed-grey lines indicate multiples of the orbital frequency and illustrate the regularity of the tidally-perturbed pulsation modes.

As demonstrated by Fig.~\ref{fig:v453:FT}, there is considerable variance remaining in the residual amplitude spectrum of V453\,Cyg, but it is not significant given our amplitude S/N $\geq 4$ criterion. The regularity amongst mode frequencies is clearly related to the orbital frequency of the system, with a common spacing of approximately 0.51\,d$^{-1}$ (i.e.\ twice the orbital frequency, which is related to the rotation frequency in a pseudo-synchronised binary). We also note that the amplitude distribution of the remaining variance is similar to the recent near-ubiquitous detection of gravity waves excited by core convection in hundreds of massive stars \citep{Bowman+19natas}. The presence of this stochastic low-frequency variability remains significant up to $\sim$30\,d$^{-1}$ above the white noise level, which we calculate to be approximately 13\,$\mu$mag.


\section{Summary and discussion}

V453\,Cyg is an extensively studied EB containing stars of mass 14 and 11\Msun, in a short-period orbit that shows eccentricity and apsidal motion. Using TESS photometry, we have discovered $\beta$\,Cep pulsations in this eclipsing system. Line profile variations in published spectroscopy indicate that these pulsation arise from the primary star. We have used the TESS light curve and published spectroscopic results to determine the physical properties of the stars to high precision. We subtracted the effects of binarity from the light curve to leave behind the pulsation signal, and then subjected this to a frequency analysis.

We found tidally-perturbed pulsations similar to those recently discovered in the Algol system U\,Gru \citep{Bowman+19apj}. Similarly, it is plausible that the tidal torque in V453\,Cyg is strong enough to perturb pulsation modes and cause regularity in the amplitude spectrum in the form of pulsation frequencies offset from orbital harmonics separated by the orbital frequency. However, there are important differences between these systems. U\,Gru is experiencing slow mass transfer, has a circular orbit, and shows $\delta$\,Sct pulsations. V453\,Cyg has undergone no mass transfer, its eccentric orbit means the tidal deformation of the stars changes continuously through each orbit, and its larger stellar masses mean the tidally-perturbed pulsation frequencies are in the $\beta$~Cep frequency regime.

It is important to obtain further spectroscopic data for V453\,Cyg to confirm that the primary star is pulsating, to check whether the secondary star is contributing any of the pulsation frequencies, and to measure the component masses to better than the 1\% precision level necessary for the highest-fidelity tests of theoretical models \citep[e.g.][]{Valle+18aa}. A new apsidal motion study would also be valuable. Armed with these empirical measurements, V453\,Cyg represents an unique opportunity for a detailed theoretical exploration of the physics that governs the structure and evolution of massive stars in short-period binary systems.

\vspace*{-5pt}


\section*{Acknowledgements}

The TESS data presented in this paper were obtained from the Mikulski Archive for Space Telescopes (MAST) at the Space Telescope Science Institute (STScI). 
STScI is operated by the Association of Universities for Research in Astronomy, Inc. 
Support to MAST for these data is provided by the NASA Office of Space Science. 
Funding for the TESS mission is provided by the NASA Explorer Program.
The research leading to these results has received funding from the European Research Council (ERC) under the European Unions Horizon 2020 research and innovation programme (grant agreement No.\ 670519: MAMSIE), from the KU Leuven Research Council (grant C16/18/005: PARADISE), from the Research Foundation Flanders (FWO) under grant agreement G0H5416N (ERC Runner Up Project), from the BELgian federal Science Policy Office (BELSPO) through PRODEX grant PLATO.

\vspace*{-5pt}


\bibliographystyle{mn_new}

\end{document}